\documentclass[a4paper, 10pt]{amsart}
\usepackage{defsWIP}

\title[Exact discrete-time optimal control of a WIP]{Structure-preserving discrete-time optimal maneuvers of a wheeled inverted pendulum}

\author[K.\ S.\ Phogat]{Karmvir Singh Phogat}
\address{Systems \& Control Engineering, IIT Bombay, Mumbai~--~400076, India}
\email[K.\ S.\ Phogat]{karmvir.p@iitb.ac.in}
\urladdr[K.\ S.\ Phogat]{\url{http://www.sc.iitb.ac.in/~karmvir.p}}

\author[R.\ Banavar]{Ravi Banavar}
\email[R.\ Banavar]{banavar@iitb.ac.in}
\urladdr[R.\ Banavar]{\url{http://www.sc.iitb.ac.in/~banavar}}

\author[D.\ Chatterjee]{Debasish Chatterjee}
\email[D.\ Chatterjee]{dchatter@iitb.ac.in}
\urladdr[D.\ Chatterjee]{\url{http://www.sc.iitb.ac.in/~chatterjee}}


\date{\today}

\begin{document}

\begin{abstract}
The Wheeled Inverted Pendulum (WIP) is a nonholonomic, underactuated mechanical system, and has been popularized commercially as the {\it Segway}. Designing optimal control laws for point-to-point state-transfer for this autonomous mechanical system, while respecting momentum and torque constraints as well as the underlying manifold, continues to pose challenging problems. In this article we present a successful effort in this direction: We employ geometric mechanics to obtain a discrete-time model of the system, followed by the synthesis of an energy-optimal control based on a discrete-time maximum principle applicable to mechanical systems whose configuration manifold is a Lie group. Moreover, we incorporate state and momentum constraints into the discrete-time control directly at the synthesis stage. The control is implemented on a WIP with parameters obtained from an existing prototype; the results are highly encouraging, as demonstrated by numerical experiments.
\end{abstract}

\maketitle
\section{Introduction}
Synthesizing discrete-time control signals for mechanical systems with state and control constraints, while preserving the underlying mechanical structure in the process of discretization, is a challenging problem. Quite often, at the control design stage, to ensure that the control objectives and constraints are satisfied, prior experience and a trial-and-error approach  are involved. Furthermore, the continuous time controller is either discretized after design or, the system dynamics is discretized to start with using a standard discretization scheme, and the control design proceeds thereafter. Both these procedures are approximations, and in the latter, one loses the mechanical nature of the system. Incorporation of constraints poses a further challenge, and this is addressed in a rather ad-hoc manner. A faithful discretization scheme followed by a computationally tractable control law synthesis, which respects multiple (state and control) constraints, is, therefore, most desirable. The variational integrator obtained from discrete mechanics \cite{dm_marsden} is a solution to the first problem, and the second issue is well addressed by posing the problem as a discrete-time optimal control problem and using a variant of the Pontryagin Maximum Principle (PMP) \cite{KarmDPMP} to obtain an open-loop solution. The optimal trajectory obtained via this open loop strategy is tracked, in general, via a close loop tracking controller. The Wheel Inverted Pendulum (WIP) is a representative mechanical system, that brings in considerable complexity like nonholonomic behaviour and underactuation in its description. This article presents our attempt in synthesizing a control law for a WIP using discrete mechanics \cite{dm_marsden} to obtain a discrete-time model (variational integrator), followed by an optimal control synthesis based on the Maximum principle \cite{KarmDPMP}. 

\begin{figure}
\centering
\includegraphics[scale=0.5]{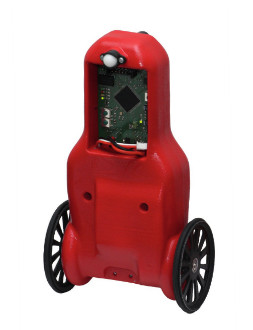}
\caption{The wheeled inverted pendulum (courtesy: Institute of Automatic Control, TU Munich)}
\label{proto}
\end{figure}

The WIP, (see Figure \ref{proto},) consists of a vertical body with two coaxial driven wheels. The system is underactuated since the number of control inputs (the drive on the wheels) are less than the number of configuration variables; in addition, the system has nonholonomic constraints that arise due to the pure rolling assumption on the wheels \cite{arXivSneha}. The WIP finds many applications, that include baggage transportation, commuting and navigation \cite{Segway2014}. The system has gained interest in the past several years due to its maneuverability and simple construction (see e.g. \cite{chan2013review,Grasser2002}). Other robotic systems based on the WIP are becoming popular as well in the robotic community for human assistance or transportation as can be seen in the works of \cite{Li2012, Nasrallah2006, Nasrallah2007, Baloh2003}, and a commercially available model $Segway$ for human transportation \cite{Segway2014}.  Several control laws have been applied to the WIP, mostly using linearized models as can be seen in \cite{blankespoor2004experimental, zhou2016robust, salerno2004control, kim2005dynamic,Li2012,vasudevan2015design}.  These control laws are limited to small tilt angle and orientation maneuvers. In \cite{salerno2003nonlinear}, controllability of the dynamics involving the rotation of the wheels and the pitch of the vertical body (pendulum) is presented, and in \cite{salerno2004control} a linear controller is designed for stabilization. In \cite{pathak2005velocity} and \cite{gans2006visual}, the authors propose a controller based on partial-feedback linearization. A nonlinear position and velocity stabilization controller using energy shaping technique has been proposed in \cite{delgado2016energy}, and a vision based tracking controller for an autonomous WIP system following a walking human has been discussed in \cite{ye2016vision}. None of these efforts account for state and control constraints at the controller design stage. Therefore, autonomous maneuvering of the system from a given initial state to a given final state needs a constrained path planning algorithm. This article complements the work exists in literature: It proposes an algorithm to generate an optimal trajectory for point-to-point state transfer satisfying state and control constraints simultaneously.

	The advantages of the proposed algorithm are three fold: 
\begin{itemize}
\item First, a variational integrator for the WIP system has been derived using discrete mechanics \cite{dm_marsden} which preserves certain system, e.g., total energy, momentum etc., unlike standard discretization schemes as Euler's step and its derivatives. 
\item Second, a discrete-time energy optimal control problem is defined on the configuration manifold and a discrete-time maximum principle \cite{KarmDPMP} is applied to arrive at a two point boundary value problem. This constrained boundary value problem is then solved using multiple shooting techniques \cite{karmmsm}. This is an indirect method for solving optimal control problems, and hence more accurate than direct optimization techniques \cite{trelat}. 
\item Third, multiple shooting methods can be implemented on a parallel architecture, thereby reducing the time to synthesize the paths. 
\end{itemize}   
    
	The article unfolds as follows: Initially, we present mechanical preliminaries of the system in \secref{sec:pre}. This is followed by a fairly detailed overview of nonholonomic systems in a geometric framework in \secref{sec:nhoverview}, and in particular, the nonholonomic connection, that bears particular relevance to the discrete Lagrange-D'Alembert-Pontryagin (LDAP) principle explained in \secref{sec:ldap}. The discrete-time variational integrator  for the WIP is derived in \secref{sec:dvarwip}, and in this context an energy-optimal control problem is posed and first-order necessary conditions for optimality are derived from \cite{KarmDPMP} in \secref{sec:optwip}. \secref{sec:reswip} is dedicated to numerical simulations and results.

\section{Preliminaries} \label{sec:pre}
\begin{figure}[ht]
% \centering
%  \psfrag{a}[l][l][0.9][0]{$\alpha$}
%  \psfrag{t}[c][l][0.9][0]{$\theta$}
%  \psfrag{x}[b][t][0.9][0]{$x$}
%  \psfrag{y}[l][l][0.9][0]{$y$}
%  \psfrag{z}[r][c][0.9][0]{$z$}
%  \psfrag{x1}[l][l][0.9][0]{$I_{W{xx}}$}
%  \psfrag{y1}[l][l][0.9][0]{$I_{W{yy}}$}
%  \psfrag{z1}[br][t][0.9][0]{$I_{W{zz}}$}
%  \psfrag{x2}[lt][l][0.9][0]{$I_{B{xx}}$}
%  \psfrag{y2}[c][rb][0.9][0]{$I_{B{yy}}$}
%  \psfrag{z2}[bl][l][0.9][0]{$I_{B{zz}}$}
%  \psfrag{wheels}[tr][c][0.9][0]{$r_w, m_w, I_W$}
%  \psfrag{wheels1}[r][t][0.9][0]{$\tau^2, \phi_2$}
%  \psfrag{wheels2}[rl][l][0.9][0]{$\tau^1, \, \phi_1$}
%  \psfrag{body}[l][l][0.9][0]{$m_b, I_B, b $}
%  \psfrag{dd}[l][l][0.9][0]{$d_w$}
%  \includegraphics[width=0.75\textwidth]{WIP_Parameters.eps}
%  latexmk -pdf --shell-escape OptimalControlWIP.tex 
% Remove "off" in 'defsWIP.sty' file.
\includegraphics[width=0.75\textwidth]{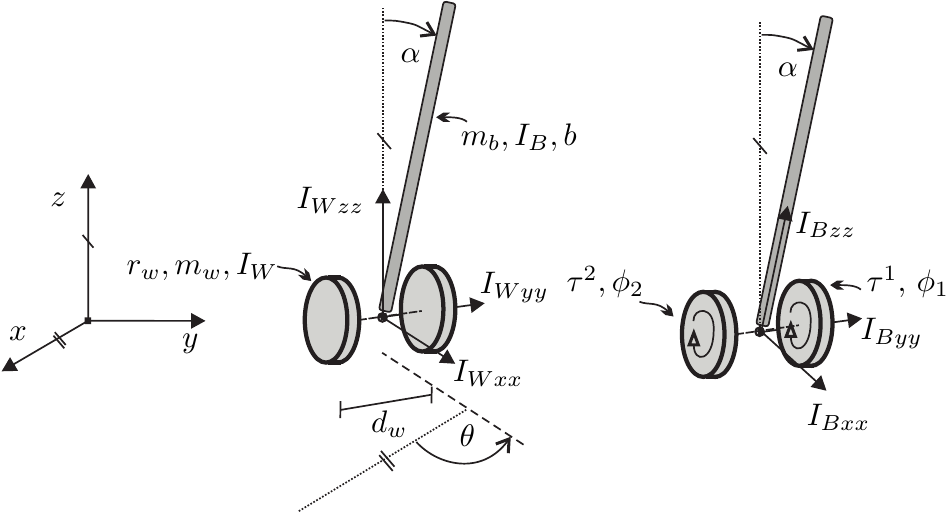}
\caption{A schematic for analysis of WIP}
\label{fig:Skizze2}
\end{figure}
The WIP, consists of a body of mass \m{\bm} mounted on wheels of radius \m{\wrd} and at a height \m{b} from the wheel rotation axis. A pair of wheels of mass \m{\wm} each, are mounted on the base of the body with a distance \m{2 \wdt} between them, and these wheels are able to rotate independently. The actuating mechanisms of the system,  typically motors, are fitted on the body to rotate the individual wheels and generate the tilting motion in the system. For these type of systems, one of the control objectives is to stabilize the body in the upward position via back and forth motion of the system.

The configuration variables of the system are: 
    \begin{itemize}
    \item \m{(x,y) \in \R^2}: the coordinates of the origin of the body-fixed frame in the horizontal plane of the inertial frame, with \m{x} as the direction along the natural rolling motion; 
    \item \m{\ha \in \s}: the heading angle (angle of the wheel rotation axis with the \m{x}-axis or the \m{y}-axis in the inertial frame),
    \item \m{\ta \in \s}: the tilt angle of the body (angle of the body \m{z}-axis with the horizontal plane in the inertial frame),  
    \item \m{\wa_1 \in \s } and \m{\wa_2 \in \s }: the relative rotation of individual wheels w.r.t. the body-fixed frame about the wheel rotation axis.
    \end{itemize}
Based on this choice, the configuration space \m{Q} of the system is \m{\left( \SE{2} \times \s \times \s \times \s \right),}   
with \m{q \Let \left( x, y, \ha, \ta, \wa_1,\wa_2 \right) \in Q.} 

The system is subject to nonholonomic constraints that arise due to no-slip conditions on the wheels, i.e., no lateral sliding and pure rotation without slipping. Let \m{\R \ni t \mapsto q(t) \in Q} denotes a system trajectory. 
%
% \textcolor{red}{Karmvir: please change to the $v$ notation below to ensure consistency later. }
%
The constraints are given by the following set of equations:
\begin{equation}
\begin{cases}
\begin{aligned}
\dot{x}_L(t) \cos \ha(t) + \dot{y}_L(t) \sin \ha(t) &= \wrd \dot{\wa}_1(t), \\
\dot{x}_R(t) \cos \ha(t) + \dot{y}_R(t) \sin \ha(t) &= \wrd \dot{\wa}_2(t), \\
-\dot{x}(t) \sin \ha(t) + \dot{y}(t) \cos \ha(t) &= 0,
\end{aligned}
\end{cases}
\end{equation}	
where \m{x_L(t)\Let x(t)-d \sin \ha(t), y_L(t) \Let x(t)+d \cos \ha(t), x_R(t) \Let x(t)+d \sin \ha(t)} and \m{y_R(t) \Let x(t)-d \cos \ha(t).}
These constraints can be further compressed to the form
\begin{equation}\label{eq:nh}
\begin{aligned}
\dot{x}(t)- \wrd \cos \ha(t) \left( \dot{\wa}_1(t) + \dot{\wa}_2(t) \right)=0,\\
\dot{y}(t)- \wrd \sin \ha(t) \left( \dot{\wa}_1(t) + \dot{\wa}_2(t) \right)=0,\\
\dot{\ha}(t) - \frac{\wrd}{\wdt} \left( \dot{\wa}_2(t) - \dot{\wa}_1(t) \right) = 0.
\end{aligned}
\end{equation}
We shall derive the discrete-time dynamics of WIP satisfying nonholonomic constraints \eqref{eq:nh} in the sequel. However, before proceeding to the discrete-time modeling of the system, we provide a brief overview of nonholonomic systems and the associated mechanical connection, in both continuous and discrete-time.

\section{Nonholonomic systems: an overview}  \label{sec:nhoverview}
In this section we will focus on the concepts which are crucial in deriving discrete-time variational integrators for nonholonomic systems. First, we give a brief introduction to constrained distributions and constrained Lagrangians, followed by the nonholonomic connection and its local form. A wealth of information about the geometry of nonholonomic systems may be found in \cite{ref:Blo-15,kobilarov2010geometric, ostrowski1996mechanics}.

\subsection{Lagrangian and constrained distribution}\label{sec:lag}
	Let \m{Q} be the configuration space of a nonholonomic mechanical system. Suppose 
\[ \lieg \times Q \ni \left(\bar{g},q\right) \mapsto \Phi_{\bar{g}} (q) \in Q \]
is a group action of a Lie group \m{\lieg} on the manifold \m{Q.}
Then the space of symmetries at a given configuration \m{q \in Q} is the orbit of \m{\lieg}: 
\[\go(q) \Let \{\left. \Phi_{\bar{g}} \left(q\right) \;\right|\; \bar{g} \in \lieg \},\]
and it is a submanifold \cite[p.\ 107]{ref:Blo-15} of \m{Q}. Let \m{\liea} be the Lie algebra associated with the Lie group \m{\lieg} and 
\[ \xi_{Q}(q) \Let \left.\frac{d}{d\epsilon}\right|_{\epsilon=0} \Phi_{\exponential^{\epsilon \xi}}(q) \]
is the infinitesimal generator of \m{\xi \in \liea.}  Then the tangent space of the orbit at a point \m{q} is given as
\[ T_{q} \go (q) = \left\{\xi_{Q}(q) \; | \; \xi \in \liea \right\}.\]

Let 
    \[TQ \ni \left(q,v_q\right) \mapsto L \left(q,v_q\right) \in \R \] 
    be the Lagrangian of the nonholonomic system with a regular distribution\footnote{A smooth distribution \m{\mathcal{D}} on a manifold \m{Q} is a smooth assignment of subspaces \m{\mathcal{D}_q \subset T_q Q} at each \m{q \in Q.} A distribution is said to be \textit{regular} \cite[p.\ 96]{ref:Blo-15} on \m{Q} if there exists an integer \m{d} such that \m{\text{dim}(\mathcal{D}_q) = d} for all \m{q \in Q.}} \m{\mathcal{D}} satisfying the nonholonomic constraints. 

The following assumptions, standard in the literature \cite{ostrowski1996mechanics,kobilarov2010geometric,ref:Blo-15}, are considered throughout the article:
\begin{enumerate}[label=(A-\roman*), leftmargin=*, widest=b, align=left]
	\item \label{asm:1} For each \m{q \in Q,\; T_{q}Q = \mathcal{D}_q+T_{q} \go (q).} 
	\item \label{asm:2} The Lagrangian \m{L} is invariant under the group action \m{\Phi,} i.e., 
\[L(q,v_q)= L\left(\Phi_{\bar{g}}\left(q\right),T_q\Phi_{\bar{g}}\left(v_q\right)\right) \text{\;for all\;} \bar{g} \in \lieg \text{\;and\;} q \in Q.\]  
  	\item \label{asm:3} The distribution \m{\mathcal{D}} is invariant under the group action, i.e., the subspace \m{\mathcal{D}_q \subset T_q Q } is translated under the tangent lift of the group action to the subspace \m{\mathcal{D}_{\Phi_{\bar{g}}\left(q\right)} \subset T_{\Phi_{\bar{g}}\left(q\right)} Q} for all \m{\bar{g} \in \lieg} and \m{q \in Q.} 
\end{enumerate}
Assumption \ref{asm:2} is the key property for defining the Lagrangian on the reduced space. \ref{asm:3} is necessary to define the constrained reduced Lagrangian and local form of the nonholonomic connection that is discussed in \secref{sec:nh}. 

	Let a principal fiber bundle \m{Q \Let \lieg \times M} be the configuration space of a mechanical system with \m{\lieg} as a Lie group, and \m{M} as a manifold that defines the shape space or the base manifold. Let \m{q \Let \left(g,s\right)} be a configuration on the manifold \m{\lieg \times M.} Then the reduced Lagrangian is defined as
\begin{align}\label{eq:rlag}
TM \times \liea \ni \left(s,v_{s},\xi\right) \mapsto \rlag\left(s,v_s,\xi\right) \Let L\left( \left(e,s\right), \left(T_{g}\Phi_{g^{-1}} \left(v_g\right), v_s\right) \right) \in \R,
\end{align} 
where \[ \xi = T_{g}\Phi_{g^{-1}} \left(v_g\right) \in \liea. \]
The momentum map \m{J : TQ \rightarrow \liea^*} corresponding to the \m{\lieg}-invariant Lagrangian is defined as
\[
\ip{J\left(q,v_q\right)}{\xi} = \ip{\frac{\partial L }{\partial v_q }\left( q,v_q \right)}{\xi_{Q}(q)} \quad \text{for all\;} \xi \in \liea,
\]
where \m{\ip{\cdot}{\cdot}} denotes the duality product for the pair\m{(g^*,g)} given by \m{g^*\times g \ni (l,v)\Let\ip{l}{v}=l(v) \in \R.}
The momentum map \m{J^b : TM\times \liea \rightarrow \liea^*} can be defined in body coordinates for an invariant Lagrangian in terms of the spatial momentum map \m{J} as
\begin{align}\label{eq:nhmm}
J^b \left(s,v_s,\xi\right) \Let \ad{g}^* \left( J \left(g,s, v_g,v_s\right) \right),
\end{align}
where \m{\ad{g}} is the adjoint action of the Lie group \m{\lieg} on the Lie algebra \m{\liea.}
	Let \m{\mathcal{V}_q} be the space of tangent vectors parallel to the symmetric directions, (i.e., vertical space), \m{\mathcal{D}_q} be the space of velocities satisfying the nonholonomic constraints at a given configuration \m{q}, \m{\mathcal{S}_q} be the space of symmetric directions satisfying nonholonomic constraints \eqref{eq:nh}, and \m{\mathcal{H}_q} be a space of tangent vectors satisfying nonholonomic constraints but not aligned with the symmetric directions. Then these subspaces of \m{T_q Q} are identified as \cite{kobilarov2010geometric,ostrowski1996mechanics}
\[ \mathcal{V}_q = T_{q} \go (q), \quad \mathcal{S}_q = \mathcal{V}_q \cap \mathcal{D}_q, \quad \mathcal{D}_q = \mathcal{S}_q \oplus \mathcal{H}_q.  \]

We are now ready to define the nonholonomic connection  and its local form. 

\subsection{Nonholonomic connection}\label{sec:nh}
\begin{definition}
A principal connection \m{\nc: T Q \rightarrow \liea} is a Lie algebra valued one form that is linear on each subspace and satisfies the following conditions:
\begin{enumerate}
\item \m{\nc(q) \cdot \xi_{Q}(q) = \xi, \quad \xi \in \liea,} and \m{q \in Q,}
\item \m{\nc} is equivariant: 
\[
\nc\left(\Phi_g\left(q\right)\right) \cdot T_q  \Phi_g \left(v_q\right) = \ad{g} \left( \nc (q) \cdot v_q \right) \text{\;for all\;} v_q \in T_q Q \text{\;and\;} g \in \lieg,
\] 
 where \m{\Phi_g} denotes the group action of \m{\lieg} on \m{Q} and \m{\ad{g}} denotes the adjoint action of \m{\lieg} on \m{\liea.} 
\end{enumerate}
\end{definition}

The principal connection determines a unique Lie algebra element corresponding to a tangent vector \m{v_q \in T_{q} Q.} For a given vertical space \m{\mathcal{V}_q} and a horizontal space \m{\mathcal{H}_q}, a vector \m{v_q \in T_{q} Q} can be uniquely represented as \m{v_q = \text{ver}\left(v_q\right) + \text{hor}\left(v_q\right),} where \m{\text{ver}\left(v_q\right) \in \mathcal{V}_q} and \m{\text{hor}\left(v_q\right) \in \mathcal{H}_q}. By the definition of the principal connection, 
\[ \nc (q) \cdot \text{ver}\left(v_q\right) = \xi,\]
where \m{\xi \in \liea} is the unique Lie algebra element associated with the vertical component \m{\text{ver}\left(v_q\right)}, i.e.,  \m{\text{ver}\left(v_q\right) = \xi_{Q}(q) \in T_q Q} for some \m{\xi \in \liea.}  
Consequently, the connection evaluates to zero on the horizontal component \m{\text{hor}\left(v_q\right)}, i.e.,
\[\nc (q) \cdot \text{hor}\left(v_q\right) = 0.\] 

	In the case of a principle fiber bundle \m{Q=\lieg \times M}, the principle connection admits a local form \m{\lf: TM \rightarrow \liea} such that the principle connection in terms of the local form is given by \cite{bloch1996nonholonomic}
\[ \nc (q) \cdot v_q = \ad{g}\left(g^{-1}(v_g) + \lf\left(s\right) v_s \right) \quad \text{for all\;} q\Let \left(g,s\right) \in Q \text{\; and \;} v_q\Let \left(v_g,v_s \right) \in T_{q}Q,\]
where \m{g^{-1}(v_g)} is the tangent lift of the left action of \m{g^{-1}} on \m{g \in \lieg}, and \m{\ad{g}: \liea \rightarrow \liea } is given by
\[ \ad{g}(\xi) \Let \left.\frac{d}{d\epsilon}\right|_{\epsilon = 0} g\exponential^{\epsilon\xi}g^{-1} \quad \text{for all\;} \xi \in \liea. \]

	For mechanical systems evolving on principle fiber bundles, in general, the base space \m{M} corresponds to the set of configurations which are directly controlled by the control forces and hence a path on the base space can be followed by applying these forces. A path on the fiber space \m{\lieg} is constructed by fiber velocities at given fiber configurations. These fiber velocities are uniquely related to nonholonomic momentum and base velocities via a nonholonomic connection. 
    Let us choose a vector subspace \m{\mathcal{U}_q \subset \mathcal{V}_q} such that 
    \[\mathcal{V}_q = \mathcal{S}_q \oplus \mathcal{U}_q, \]
 where \m{\mathcal{D}} is the distribution satisfying nonholonomic constraints and \m{\mathcal{S}} is a distribution consists of the symmetric horizontal directions.
 
\begin{definition}[{{\cite[Definition 6.2 on p.\ 38]{bloch1996nonholonomic}}}]
Consider that the Assumption \ref{asm:1} holds. Then the nonholonomic connection \m{A^{\text{nhc}} : TQ \rightarrow \mathcal{V}} is a vertical valued one form whose horizontal space at \m{q \in Q} is the orthogonal complement of the subspace \m{\mathcal{S}_q} in \m{\mathcal{D}_q} and satisfies the following:
\[A^{\text{nhc}} \Let A^{\text{kin}} + A^{\text{sym}}, \]
where  \m{A^{\text{kin}}: TQ \rightarrow \mathcal{U}} is the kinematic connection  enforcing nonholonomic constraints and \m{A^{\text{sym}}: TQ \rightarrow \mathcal{S}} is the mechanical connection corresponding to symmetries in the constrained direction.     
\end{definition}	
        
The kinematic connection \m{A^{\text{kin}}} and the mechanical connection \m{A^{\text{sym}}} satisfy the following conditions:

\[
A^{\text{kin}} (q) \cdot v_q = 0 \quad \text{for all\;} v_q \in \mathcal{D}_q, \quad 
A^{\text{sym}}(q) \cdot v_q = v_q\quad \text{for all\;} v_q \in \mathcal{S}_q.
\]
\begin{remark}
If the distribution \m{\mathcal{S}_q} and the horizontal distribution are invariant under the group action, then the nonholonomic connection is a principal connection.
\end{remark}
In case the nonholonomic connection is a principal connection, the connection is represented as
\[
\ad{g}\left( g^{-1}(v_g) + \lf\left(s\right) v_s\right) = \ad{g} \left(\Omega \right),
\]
where 
\[ 
\Omega \in \gs_s \Let \left\{ \xi \in \liea \;|\; \xi_{Q}(q) \in \mathcal{S}_q \right\}
\]
is the \emph{locked angular velocity} i.e. the velocity achieved by locking the joints represented by the base configuration variable.
This local form of the nonholonomic connection can be written as
\begin{align}\label{eq:clf}
g^{-1}(v_g) + \lf\left(s\right) v_s = \Omega.
\end{align} 
For the principal kinematic case, i.e., \m{\mathcal{S}_q = \mathcal{D}_q \cap \mathcal{V}_q = \{0\}} for all \m{q \in Q,} the local form of the nonholonomic connection \eqref{eq:clf} simplifies to
\[ 
g^{-1}(v_g) + \lf\left(s\right) v_s  =0.
\]
Therefore, for a smooth curve \m{\R \ni t \mapsto \left(g\left(t\right),s\left(t\right)\right) \in \lieg \times M,} the group motion can be constructed by the nonholonomic connection for a given base trajectory as
\[
\dot{g}(t) = - g(t) \lf \left(s(t)\right) \dot{s}(t).
\]

The constrained reduced Lagrangian \m{\crlag: TM \times \gs \rightarrow \R } can be defined using the local form of the nonholonomic connection as
\begin{align}
\crlag \left(s,v,\Omega\right) \Let \rlag \left( s, v,\Omega - \lf (s) v \right).
\end{align}

    In the next section, we will discuss a unified approach to derive discrete-time variational integrators using the reduced Lagrange-D'Alembert-Pontryagin nonholonomic principle \cite{kobilarov2010geometric}.

\section{Discrete-time variational integrator} \label{sec:ldap}
	In this section we present the discrete-time analogue of the geometric objects discussed in \secref{sec:nh}. Then we state the discrete-time reduced Lagrange-D'Alembert-Pontryagin nonholonomic principle \cite{kobilarov2010geometric} and discuss the variational integrators derived for nonholonomic systems by applying this principle.

	The local form of the nonholonomic connection \eqref{eq:clf} can be represented in discrete-time  as
\begin{align*}
\xi_k + \lf(s_k) v_k = \Omega_k
\end{align*}
where \m{v_k = (s_{k+1}-s_k)/h, \; g_{k}^{-1}g_{k+1} = \varphi\left(h \xi_k\right), \;h} is the time difference between two consecutive configurations, i.e., \m{t_{k+1}-t_k} such that \m{s_k\Let s(t_k), } and the map \m{\varphi:\liea \rightarrow \lieg} represents the \emph{difference} between two system configurations defined via Lie group elements by a unique element in its Lie algebra. In most of the cases, \m{\varphi} is taken to be the exponential map \m{\exponential:\liea \rightarrow \lieg} that is a diffeomorphism in the neighborhood \m{\mathcal{O}_e \subset \lieg } of the group identity \m{e \in \lieg} \cite[p.\ 256]{abraham}. The map \m{e} serves the purpose of \m{\varphi} because the consecutive group configurations \m{g_k} and \m{g_{k+1}} do not differ by a large value, i.e., \m{g_k^{-1}g_{k+1} \in \mathcal{O}_e \subset \lieg} for any discrete-time instant \m{k.} Similarly, the discrete-time body momentum map \m{J^b:TM \times \liea \rightarrow \liea^*} is defined in terms of the spatial momentum map \m{J} \eqref{eq:nhmm} as
    \[ J^b \left(s_k,v_k,\xi_k\right) \Let \ad{g_k}^* \left( J \left(g_k, s_k, v_{gk}, v_k \right) \right)\]
    where \m{ \ad{g_k}^*: \liea^* \rightarrow \liea^*} is the map that translates momentum vectors from the spatial frame to the body frame and \m{\xi = g^{-1}(v_q)} for \m{v_g \in T_g \lieg.}
    
	Let us define a discrete path 
\begin{align}
\left(s,v,p,g,\Omega,\mu \right) : \{t_k\}_{k=0}^N \rightarrow TM \times T^*M \times \lieg \times \gs \times \liea^*,  
\end{align}
on the reduced space that satisfies the following constraints
\[s_{k+1} - s_{k} = h v_{k}, \quad g_{k+1} = g_{k} \varphi (h \xi_k), \]
where \m{h \xi_k = \Omega_k - \lf(s_{k+\beta}) v_k}, \m{s_{k+\beta} \Let \beta s_{k+1} + (1-\beta)s_{k}} for a chosen \m{\beta \in [0,1]}.

Similarly, the discrete control force \m{\tau : \{t_k\}_{k=0}^N \rightarrow T^{*}M} is an approximation of the continuous-time force controlling the shape of the dynamics.
    
    \begin{definition}{The Discrete Reduced LDAP Principle}
    \begin{equation}
    \begin{aligned}
    \delta \sum_{k=0}^{N-1} & h \Big[\crlag (s_{k+\beta},v_k,\Omega_k )  + \ip{\mu_k}{\varphi^{-1}\left(g_k^{-1}g_{k+1} \right)/h + \lf\left(s_{k+\beta}\right) v_k - \Omega_k} \\
   & + \ip{p_k}{\left(s_{k+1}-s_k\right)/h - v_k} \Big] + \delta \sum_{k=0}^{N-1} h \Big[\ip{\tau_{k+\beta}}{s_{k+\beta}} \Big] = 0,
    \end{aligned}    
    \end{equation}       
subject to 
\begin{equation}
\begin{aligned}
& \text{vertical variations s.t.\;} \left( g_k^{-1} \delta g_k,\delta s_k\right) = \left(\eta_k, 0 \right), \quad \eta_k \in \gs_{s_k} \; \text{and}, \\
& \text{horizontal variations s.t.\;} \left( g_k^{-1} \delta g_k,\delta s_k\right) = \left(-\lf(s_k) \delta s_k, \delta s_k\right).
\end{aligned}
\end{equation}
\end{definition}
The discrete reduced LDAP principle leads to the following sets of discrete-time equations:
\begin{equation}\label{eq:dnc}
\begin{aligned} 
s_{k+1} & = s_k + h v_k, \\
g_{k+1} & = g_{k} \varphi \left( h \xi_k \right),
\end{aligned}
\end{equation}
and the discrete-time momentum equation in an implicit form
\begin{align}\label{ldap:meq}
\ip{J^b (s_k,v_k,\xi_k) + J^b (s_{k-1},v_{k-1},\xi_{k-1})}{e_{b}(s_k)} = 0 \; \text{for\;} b=1,\ldots,\text{dim}(\gs_{s_k}),
\end{align}
where  \m{\{e_{b}(s_k)\} \in \gs_{s_k} } is a basis of the Lie algebra,
and the horizontal equation
\begin{align}\label{eq:heq}
& \left(\frac{\partial \rlag_{k+\beta}}{\partial v} - \frac{\partial \rlag_{k-1+\beta}}{\partial v} \right) - h \left(\beta \frac{\partial \rlag_{k-1+\beta}}{\partial s} + \left(1-\beta\right) \frac{\partial \rlag_{k+\beta}}{\partial s} \right) \\ \nonumber
& = \lf^*(s_{k}) \left( J^b (s_k,v_k,\xi_k) - J^b (s_{k-1},v_{k-1},\xi_{k-1}) \right) + h \left(\beta \tau_{k-1+\beta} + (1-\beta)\tau_{k+\beta} \right)\\
& \quad \text{for} \; k=1,\ldots,N-1, \nonumber
\end{align}
 where \m{\xi_k= \Omega_k - \nc(s_{k+\beta})v_{k}} and \m{\rlag_{k+\beta} \Let \rlag\left(s_{k+\beta},v_k,\xi_k \right)}
 is the reduced Lagrangian \eqref{eq:rlag}.
  In the next section we derive the variational integrator for the WIP mechanism.

\section{Discrete-time model of the WIP mechanism} \label{sec:dvarwip}
We now employ the tools elaborated in the previous section to derive a discrete-time model of the WIP mechanism that was presented in \secref{sec:pre}. In the notations established above, with  \m{\lieg = \SE{2}} as the Lie group, the configuration space \m{Q} of the WIP system can be written in the \emph{trivial bundle} form as 
\[ Q = \lieg \times M \Let \SE{2} \times \left( \s \times \s \times \s \right). \]
	The Lagrangian of the WIP is the total kinetic energy minus the potential energy  \cite{arXivSneha} and is given by
\begin{equation}\label{eq:lag}
\begin{aligned}
L(q,v_q)& = \cx \left(v_x^2 +v_y^2\right) + \ct v_{\ha}^2 + \ca v_{\ta}^2 \\
&\quad + \cp \left(v_{\wa_1}^2+v_{\wa_2}^2\right) + \caxa \left\{\cos\ta \cos \ha v_{\ta} v_{x} - \sin \ta \sin \ha v_{x}v_{\ha}\right\} \\
& \quad + \caxa \left\{ \sin \ta \cos \ha v_{\ha}v_{y} + \cos \ta \sin \ha v_{\ta} v_{y} - \mathrm{g} \cos \ta \right\}
\end{aligned}
\end{equation}
where 
\[I_{\ha}(\ta) = 2 I_{Wzz} + I_{Bzz} \cos^2 \ta + 2 m_w d^2 + \left(I_{Bxx} + m_b b^2 \right) \sin^2 \ta. \]
We now enlist a few geometric objects associated with the WIP:
\begin{itemize}[leftmargin=*]
\item \textit{Group action} (See \secref{sec:nhoverview}): The map \m{\Phi: \lieg \times Q \rightarrow Q} is the group action of the Lie group \m{\lieg} on the manifold \m{Q} and is defined in coordinates as
\begin{equation}
\Phi_{\bar{g}} (q) = \left(X+  x \cos \Theta - y \sin \Theta , Y + x \sin \Theta + y \cos \Theta, \Theta + \ha, \ta, \wa_1, \wa_2\right)
\end{equation}
\item \textit{Tangent lift} (See \secref{sec:nhoverview}): The tangent lift of the group action \m{\Phi_{\bar{g}}} is defined in coordinates as
\begin{equation}
T_{q} \Phi_{\bar{g}} (v_{q}) = \left(v_{x} \cos \Theta - v_{y} \sin \Theta, v_{x} \sin \Theta + v_{y} \cos \Theta, v_{\ha},v_\ta,v_{\wa_1}, v_{\wa_2}\right)
\end{equation}
where 
\begin{align*}
& \bar{g}\Let \left( X,Y,\Theta\right),\quad
q=\left(g,s\right) = \left(x,y,\ha,\ta,\wa_1,\wa_2 \right), \quad
v_{q}= \left(v_{g},v \right) = \left(v_{x},v_{y},v_{\ha}, v_\ta,v_{\wa_1}, v_{\wa_2} \right).
\end{align*}
\item \textit{Reduced Lagrangian} (See \secref{sec:nhoverview}): Let 
\[
\liea \times TM \ni \left(\left(e,s\right), \left(\xi,v\right)\right) = \left(\Phi_{g^{-1}}(q), T_{q}\Phi_{g^{-1}} \left(v_{q}\right)\right)
\] 
be the point on the reduced space, where 
\begin{align} \label{eq:tlfa}
 \left(\xi, v \right) = \left( \left( v_{x}\cos \ha + v_{y} \sin \ha, -v_{x} \sin \ha + v_{y} \cos \ha, v_{\ha}\right), \left( v_\ta, v_{\wa_1}, v_{\wa_2} \right) \right).  
\end{align}
Then the reduced Lagrangian is defined by
\begin{align} 
\rlag(s,v,\xi) & \Let L\left(\Phi_{g^{-1}}(q),T_{q}\Phi_{g^{-1}}\left(v_{q}\right)\right)\\ \nonumber
& = \cx \left(\xi_{1}^2 + \xi_{2}^{2} \right) + \ct \xi_{3}^2 + \ca \vta^2 \\ \nonumber
& \quad + \cp \left(\vwa{1}^2+\vwa{2}^2\right) + \caxa \sin \ta \xi_2 \xi_3 \\ \nonumber
& \quad + \caxa \cos \ta \xi_1 \vta - \caxa \mathrm{g} \cos \ta,
 \end{align}
where \m{v=\left( \vta, \vwa{1},\vwa{2} \right)} and \m{s=\left( \ta, \wa_{1},\wa_{2} \right).} 
\item \textit{Vertical Space} (See \secref{sec:nhoverview}): The vertical space for the system is given by
\begin{align*}
\mathcal{V}_q &= \left\{\left.\left.\frac{d}{d\epsilon}\right|_{\epsilon=0} \left( \gamma(\epsilon),s \right) \;  \right|\; \gamma(t) = \exponential^{t \xi}g \in \lieg, \;\gamma(0)=g, \; \dot{\gamma}(0)=v_g, \;\xi \in \liea, \; s \in M \right\},\\
&= \big\{\left(v_{g}, 0 \right) \in T_{g}G \times T_{s}M  \big\}.
\end{align*}
For a given local representation of the tangent vectors \m{v_{g} \Let \left( v_{x},v_{y},v_{\theta} \right) \in T_g G}, the local basis of the vertical space \m{\mathcal{V}_q} is given by
\[ \mathcal{V}_q = \text{span} \left\{\pr{x},\pr{y}, \pr{\ha} \right\}. \]

\item  \textit{Constrained distribution}: The distribution \m{\mathcal{D}} satisfying nonholonomic constraints \eqref{eq:nh} is called the constrained distribution. The local generator (a collection of linearly independent vector fields spanning the distribution) of the constrained distribution \m{\mathcal{D}_q} satisfying the nonholonomic constraints \eqref{eq:nh} is given by
\[\mathcal{D}_q = \text{span} \{\mathcal{X}_1,\mathcal{X}_2,\mathcal{X}_3\} \]
where
\begin{align*}
\mathcal{X}_1 &= \cos \ha \pr{x} - \sin \ha \pr{y} + \frac{1}{\wrd} \pr{\wa_1} + \frac{1}{\wrd} \pr{\wa_2}, \\
\mathcal{X}_2 &= \pr{\ta}, \\
\mathcal{X}_3 &= \pr{\ha} + \frac{\wdt}{\wrd} \pr{\wa_1} - \frac{\wdt}{\wrd} \pr{\wa_2}. 
\end{align*}
\item  \textit{Body momentum map}: The momentum map \m{J: TM\times \liea \rightarrow \liea} is defined by
\[ \ip{J\left(s,g,v,v_{g} \right)}{\xi} \Let \ip{\frac{\partial L}{\partial v_{q}}\left(q,v_{q}\right)}{\xi_{Q}(q)}\]
where 
\[ T_{q}Q \ni \xi_{Q}(q) \Let \left.\frac{d}{d\epsilon}\right|_{\epsilon=0} \Phi_{\exponential^{\epsilon\xi}}(q), \quad q \in Q.\]
The vector field \m{\xi_{Q}} in local coordinates is given as
\[ \xi_{Q}(x,y,\theta,\alpha,\phi_1,\phi_2) \Let \left(\xi_1 - y \xi_3, \xi_2 + x \xi_3, \xi_3, 0,0,0 \right).\]
The discrete-time body momentum map \m{J^b: TM\times \liea \rightarrow \liea} is defined by

\begin{align}
J^b \left(s_k,v_k,\xi_k \right) & \Let \ad{g_k}^* \left(J\left(s_k,g_k,v_k,v_{g k} \right) \right)\\ \nonumber
& = \begin{pmatrix}
\cxa \xi_{1k} + \caxa \cos \ta_{k} v_{\ta k} \\
\cxa \xi_{2k} + \caxa \sin \ta_{k} \xi_{3k} \\
\ctka \xi_{3k} + \caxa \sin \ta_{k} \xi_{2k}
\end{pmatrix},
\end{align}
where
\begin{align*}
\ad{g_k}^* \left(\mu\right) \Let \begin{pmatrix}
\cos \ha_k & \sin \ha_k & 0 \\ - \sin \ha_k & \cos \ha_k & 0 \\
y_k & -x_k & 1
\end{pmatrix} \begin{pmatrix}
\mu^1 \\ \mu^2 \\ \mu^3
\end{pmatrix}.
\end{align*}

\item  \textit{Local nonholonomic connection}:  We establish that under the choice of the group action \m{\Phi}, this system falls in the category of a \textit{principal kinematic} case, i.e., \m{\mathcal{V}_q \cap \mathcal{D}_q = \{0\}.}
Thus, it can be seen that
\[\mathcal{S}_q = \mathcal{V}_q \cap \mathcal{D}_q = \{0\}. \]
The class of systems for which \m{\mathcal{S}_q= \{0\}} falls into a spacial category in which the tangential directions along symmetry are independent of the constrained tangential directions \cite{kobilarov2010geometric, arXivSneha}. This is the \emph{principal kinematic}, case in which there are no momentum equations \eqref{ldap:meq}. We know from \eqref{eq:tlfa} that
\begin{align}\label{eq:xi}
g^{-1}\left(v_{g}\right) = \xi = \left( v_{x}\cos \ha + v_{y} \sin \ha, -v_{x} \sin \ha + v_{y} \cos \ha, v_{\ha}\right).  
\end{align}
With the adopted current convention, the \m{\dot{q}(t) \in T_{q}Q} in \secref{sec:pre} is defined by
\[\dot{q}(t) = \left(\dot{x}(t),\dot{y}(t),\dot{\theta}(t), \dot{\alpha}(t), \dot{\phi}_1(t), \dot{\phi}_2(t) \right) \Let \left(v_{x}, v_{y}, v_{\theta}, v_{\alpha}, v_{\phi_1}, v_{\phi_2}\right), \]
and further, substituting the value of \m{v_{x}, v_{y}, v_{\ha}} from \eqref{eq:nh} into \eqref{eq:xi} we obtain the local form of the nonholonomic connection as
\[ \xi + \lf(s) v = 0,\]
where 
\begin{align*}
\lf(s) =  \begin{pmatrix}
			0 & -\wrd & -\wrd \\
            0 & 0 & 0 \\
            0 & \frac{\wrd}{\wdt} & -\frac{\wrd}{\wdt}
			\end{pmatrix},  \quad 
v = \begin{pmatrix}
			v_\ta \\
            v_{\wa_1} \\
            v_{\wa_2} 
			\end{pmatrix}.
\end{align*}
\end{itemize}

\subsection{Discrete-time dynamics of the WIP} \label{ssec:dvarwip}

	We are now in a position to define the discrete-time dynamics \eqref{eq:dnc}-\eqref{eq:heq} on the configuration manifold \m{\lieg \times M} as
\begin{subequations}\label{eq:viwip}
\begin{align}
& g_{k+1} = g_{k} \exponential^{- h \lf v_{k}}, \label{eq:vigroup} \\
& s_{k+1} = s_{k} + h v_{k} , \label{eq:vibasep}\\
& \mathbb{M}\left(\ta_{k+1}\right) v_{k+1} - h \mathbb{C}\left(\ta_{k+1}, v_{k+1}\right) = \mathbb{M}\left(\ta_{k}\right) v_{k} + h \tau_{k}, \label{eq:vibasev}
\end{align}
\end{subequations}
where  \m{h>0} is the step length (its selection procedure is discussed at length in \secref{sec:ldap}),
\begin{align*}
& s \Let \begin{pmatrix}
\ta & \wa_{1} & \wa_{2}
\end{pmatrix}^\top, \quad 
v \Let \begin{pmatrix}
\vta & \vwa{1} & \vwa{2}
\end{pmatrix}^\top, \quad
\tau \Let \begin{pmatrix}
0 & \tau^{1} & \tau^{2}
\end{pmatrix}^\top,\\
& \tau^{j} \text{\; is the control torque applied along the \m{j}th wheel rotation axis,} \\
& \mathbb{M}\left(\ta\right)  \Let \begin{pmatrix}
\caa & \caxa \wrd \cos \ta  & \caxa \wrd \cos \ta \\
\caxa \wrd \cos \ta &  \mathbb{H}(\ta) +\cpa & \mathbb{K}(\ta) \\
\caxa \wrd \cos \ta & \mathbb{K}(\ta) & \mathbb{H}(\ta) + \cpa
\end{pmatrix}, \\
& 
\mathbb{K}(\ta)  \Let \wrd^2 \left(\cxa - \frac{\cta}{2\wdt^2} \right) ,  \quad 
\mathbb{H}(\ta)  \Let \wrd^2 \left( \cxa + \frac{\cta}{2\wdt^2} \right) , \\
&
\mathbb{C}\left(\ta, v \right) \Let \Big(\frac{\wrd^2}{2 \wdt^2}\left( I_{Bxx}-I_{Bzz}+m_{b} b^2 \right) \sin 2\ta \left(\vwa{1}-\vwa{2} \right)^2 \\
 & \phantom{\mathbb{C}\left(\ta, v\right) = \Big(} - \caxa \wrd \sin \ta\; \vta \left( \vwa{2} + \vwa{1}\right) + \caxa \mathrm{g} \sin \ta , 0,0 \Big)^\top, \\
& \cta \Let 2 I_{Wzz} + I_{Bzz} \cos^2 \ta + 2 m_w d^2 + \left(I_{Bxx} + m_b b^2 \right) \sin^2 \ta.
\end{align*}
A few comments are in order here. \eqref{eq:vigroup} governs the update of the system orientation and translation in the \m{x-y} plane for a motion in the base space \m{M}, and \eqref{eq:vibasep} denotes the update of the tilt and wheel angles. \eqref{eq:vibasev} expresses the dynamics with $\mathbb{M}$ matrix denoting the dependence of the inertia matrix (or the Riemannian metric) on the tilt angle $\alpha$, and the $\mathbb{C}$ matrix accounts for the Coriolis terms and the gravitational force. 
Let \m{\left(g_k,s_k,v_k\right)}  be the states of the system at a discrete instant \m{k}. Then the states at \m{(k+1)}th instant are computed in the following manner:
\begin{enumerate}
\item  Compute the group update \m{g_{k+1}} (orientation and position of the system in \m{x-y} plane) using \eqref{eq:vigroup} for a given \m{g_k} and \m{v_k}.
\item Compute the base configuration update \m{s_{k+1}} (wheels and tilt angles) using \eqref{eq:vibasep} for a given \m{s_k} and \m{v_k}.
\item If one substitute \m{\alpha_{k+1}} from \eqref{eq:vibasep} in \eqref{eq:vibasev}, then \eqref{eq:vibasev} is an implicit form in \m{v_{k+1}} for given states \m{v_k, s_k} and control torque \m{\tau_k}. This implicit form is further solved to obtain \m{v_{k+1}} (wheels and tilt angles rates)  using Newton's root finding algorithm.
\end{enumerate}

This concludes our effort to obtain a discrete-mechanics model of the WIP. The following section defines an optimal control problem in the context of \eqref{eq:viwip}.

\section{Constrained Energy Optimal Control of WIP } \label{sec:optwip}

	\textbf{Control objective:} The optimal control problem is to generate an energy optimal trajectory (for the discrete-time variational integrator derived for the WIP in \secref{ssec:dvarwip}) to move the WIP from a given initial orientation \m{g^i \in \lieg }, initial wheel configuration and tilt angle configuration \m{\left(s^i, v^i\right) \in M \times \R^3} to a final orientation \m{g^f \in \lieg }, final wheel velocity and tilt angle configuration \m{\left(\alpha^f, v^f\right) \in \text{S}^1 \times \R^3} in \m{N} discrete time-steps subject to  the following  state and control constraints:
\begin{enumerate}
\item  \m{\abs{\alpha_k} \leq a \quad \text{for all\;} k=1,\ldots,N-1,} \;\;\;{\it (limits on the tilt angle)}
\item \m{\norm{v_k}_{\infty} \leq \nu \quad \text{for all\;} k=1,\ldots,N-1,} {\it (limits on the wheel and tilt velocities)}
\item \m{\norm{\tau_k}_{\infty} \leq \mu \quad \text{for all\;} k=0,\ldots,N-1.} {\it (limits on the control effort)}
\end{enumerate}
In summary, our optimal control problem in discrete-time is defined as
\begin{equation}
\label{eq:ocpwip}
\begin{aligned}
\minimize_{\left(\tau_k\right)_{k=0}^{N-1}} &&& \cost \left(\tau\right) \Let \sum_{k=0}^{N-1} \frac{h}{2}\ip{\tau_k}{\tau_k} \\
\text{subject to} &&&
\begin{cases} 
\begin{cases} 
g_{k+1} = g_{k} \exponential^{- h \lf v_{k}} \\
s_{k+1} = s_{k} + h v_{k} \\
z_{k+1}= \mathbb{M}\left(\ta_{k+1}\right) v_{k+1} - h \mathbb{C}\left(\ta_{k+1}, v_{k+1}\right)\\  
z_{k+1}=\mathbb{M}\left(\ta_{k}\right) v_{k} + h \tau_{k}
\end{cases} \text{for\;} k =0,\ldots,N-1,\\
\norm{\tau_k}_{\infty} \leq \mu \quad \text{for} \quad k=0,\ldots,N-1, \\
\frac{1}{2}\left(\left(v^j_k\right)^2- \nu^2\right)\leq 0 \quad \text{for\;} k =1,\ldots,N-1,\text{\;and\;} j=1,2,3,\\
\frac{1}{2}\left(\left(\alpha_k\right)^2- a^2\right)\leq 0 \quad \text{for\;} k =1,\ldots,N-1,\\
 \left(g_0,s_0,v_0\right) = \left(g^i,s^i, v^i\right),\\  
 \left(g_N,\alpha_N,v_N \right) = \left(g^f, \alpha^f, v^f \right).
 \end{cases}
\end{aligned}
\end{equation}
A set of first order necessary conditions for optimality in \eqref{eq:ocpwip} is given by \cite[Corollary 2.9]{KarmDPMP}, which yields:

	Let \m{\hat{\left(\cdot \right)} : \R^3 \rightarrow \mathfrak{se}(2)^{*} } be a vector space homeomorphism. Define the Hamiltonian for the optimal control problem \eqref{eq:ocpwip} as
\begin{align}
& \mathfrak{se}(2)^{*} \times \R^3 \times \R^3 \times \text{SE}(2) \times \R^3\times \R^3 \times \R^3 \times \R^2 \ni \left(\hat{\zeta}, \psi, \lambda,  g,s, z,v, \tau \right) \mapsto \\
& H^\eta \left(\hat{\zeta}, \psi, \lambda,  g,s, z,v, \tau \right) \Let \frac{\eta h}{2} \ip{\tau}{\tau} - \ip{\hat{\zeta}}{h\lf v} + \ip{\psi}{s+h v} + \ip{\lambda}{M v + h \tau} \in \R \nonumber 
\end{align}
If \m{\{\op{\tau}_k\}_{k=0}^{N-1}} be an optimal control that solves the problem \eqref{eq:ocpwip}, and \m{\left(\op{q}\right)_{k=0}^N} is the corresponding optimal state trajectory, then there exist an adjoint trajectory 
\m{\left\{\left(\hat{\zeta}^k,\psi^k,\lambda^k\right)\right\}_{k=0}^{N-1}} on the cotangent bundle \m{\mathfrak{se}(2)^{*}\times \R^3 \times \R^3}, a sequence of covectors \m{\left\{\left(\sigma^k,\beta^k\right)\right\}_{k=1}^{N-1} \subset \R \times \R^3,} and a scalar \m{\eta \in \{-1, 0\}}, not all zero such that:
% \left\{\left(\op{g}_k, \op{s}_{k},\op{z}_{k}, \op{v}_{k}\right) \right\}_{k=0}^{N}
\begin{enumerate}[leftmargin=*, label=(\roman*), widest=b, align=left]
\item State and adjoint system dynamics
\begin{align}\label{eq:adjwip}
& \begin{cases}
\zeta^{k-1} = \ad{\exponential^{h\lf\op{v}_k}}^*\zeta^k,\\
\begin{pmatrix}
I & \mathcal{D}_s\left(\mathbb{M}\left(\op{\ta}_{k}\right) \op{v}_k \right)^\top\\
h I & \mathbb{M}\left(\op{\ta}_{k}\right)^\top 
\end{pmatrix}
\begin{pmatrix}
\psi^k \\ \lambda^k
\end{pmatrix}
+ \begin{pmatrix}
\left(\sigma^k\op{\alpha}_k \;\;0 \;\;0 \right)^\top \\ -h \lf^\top \zeta^k
\end{pmatrix}\\
+ \begin{pmatrix}
0 \\ \beta^k \odot \op{v}_k
\end{pmatrix} 
= 
\begin{pmatrix}
I & \mathcal{D}_s\left(\mathbb{M}\left(\op{\ta}_{k}\right) \op{v}_k - h \mathbb{C}\left(\op{\ta}_{k}, \op{v}_{k}\right)\right)^\top\\
0 & \mathcal{D}_v\left(\mathbb{M}\left(\op{\ta}_{k}\right) \op{v}_k - h \mathbb{C}\left(\op{\ta}_{k}, \op{v}_{k}\right)\right)^\top
\end{pmatrix}
\begin{pmatrix}
\psi^{k-1} \\ \lambda^{k-1}
\end{pmatrix},
\end{cases}
\\ \label{eq:stwip}
& \begin{cases}
\op{g}_{k+1} = \op{g}_{k} \exponential^{-h\lf \op{v}_{k}},\\
\op{s}_{k+1} =  \op{s}_{k} + h \op{v}_{k},\\
 \mathbb{M}\left(\op{\ta}_{k+1}\right) \op{v}_{k+1} - h \mathbb{C}\left(\op{\ta}_{k+1}, \op{v}_{k+1}\right)=\mathbb{M}\left(\op{\ta}_{k}\right) \op{v}_{k} + h \op{\tau}_{k}, \\
\end{cases}
\end{align}
where \m{\beta^k \odot v_k \Let \begin{pmatrix}
\beta_1^k \odot v^1_k & \beta_2^k \odot v^2_k & \beta_3^k \odot v^3_k
\end{pmatrix}^\top} is the standard Schur product;
\item Transversality conditions
\begin{align}\label{eq:transwip}
\psi^N_2 = 0 \quad \text{and} \quad \psi^N_3 = 0;
\end{align}
\item complementary slackness conditions
\begin{equation}\label{eq:compwip}
\begin{cases}
\begin{aligned}
& \sigma^{k} \left(\op{\alpha}_{k}^2 - a^2 \right) = 0 \quad \text{for all} \quad k=1,\ldots,N-1, \\
& \beta^{k}_j \left(\left(\op{v}^j_{k}\right)^2 - \nu^2 \right) = 0 \quad \text{for all} \quad k=1,\ldots,N-1, \text{\;and\;} j=1,2,3; 
\end{aligned}
\end{cases}
\end{equation}

\item non-positivity condition
\[ \sigma^k, \beta^{k}_j  \leq 0 \quad \text{for all}\quad t = 1,\ldots,N-1, \text{\; and \;} j=1,2,3;\]

\item Hamiltonian maximization pointwise in time 
\begin{align}  \label{eq:hammax}
H^\eta \left(\hat{\zeta}^k,\psi^k,\lambda^k,\op{g}_k,\op{s}_{k}, \op{z}_{k}, \op{v}_{k}, \op{\tau}_k \right) \Let \maximize_{\norm{w}_{\infty} \leq \mu } H^\eta \left(\hat{\zeta}^k,\psi^k,\lambda^k,\op{g}_k,\op{s}_{k}, \op{z}_{k}, \op{v}_{k}, w \right).
\end{align}
\end{enumerate}
\begin{remark}
Note that since the Hamiltonian \eqref{eq:hammax} is concave in \m{\tau}, the non-positive gradient condition \m{(ii)} in \cite[Corollary 2.9]{KarmDPMP}, in this case leads to the maximization of the Hamiltonian pointwise in time. 
\end{remark}
It follows that our optimal actions satisfy
\begin{align*}
\op{\tau}_{k} & = \underset{\norm{w}_{\infty} \leq \mu}{\argmax} \; H^\eta \left(\hat{\zeta}^k,\psi^k,\lambda^k,\op{g}_k,\op{s}_{k}, \op{z}_{k},\op{v}_{k}, w \right),\\
& = \underset{\norm{w}_{\infty} \leq \mu}{\argmax} \; \frac{\eta h}{2} \ip{w}{w} + h \left(\lambda_2^k w^1 + \lambda_3^k w^2 \right).
\end{align*}
In the case of a normal extremal, i.e., \m{\eta=1,} 
\begin{align}
\op{\tau}^j_{k} = \begin{cases}
\mu \quad &\text{if} \; \lambda^k_{j+1} \geq \mu, \\
 -\mu \quad &\text{if} \; \lambda^k_{j+1} \leq -\mu, \\
 \lambda^k_{j+1} \quad & \text{elsewhere}.
\end{cases}
\end{align}
In the case of an abnormal extremal, i.e., \m{\eta = 0},
\begin{align}
\op{\tau}^j_{k} \in \begin{cases}
\{\mu\} \quad &\text{if} \; \lambda^k_{j+1} >0, \\
[-\mu, \mu] \quad &\text{if} \; \lambda^k_{j+1} =0,\\ 
\{-\mu\} \quad &\text{if} \; \lambda^k_{j+1} <0. 
\end{cases}
\end{align}

\begin{remark}
when \m{\eta=1,} the optimal control \m{\op{\tau}_{k}} is the saturation function of the co-state vector \m{\left(\lambda_2^k, \lambda_3^k \right)^\top}, and when \m{\eta = 0}, the control is bang-bang.
\end{remark}

The constrained boundary value problem \eqref{eq:adjwip}-\eqref{eq:stwip} subject to boundary conditions \m{ \left(g_0,s_0,v_0 \right) = \left(g^{i}, s^{i},v^{i}\right),  \left( g_N, \alpha_N, v_N \right) = \left(g^{f}, \alpha^{f},v^f\right),} the transversality conditions \eqref{eq:transwip}, the complementary slackness conditions \eqref{eq:compwip} and the state constraints \m{\norm{v_{k}} \leq \nu, \abs{\alpha_{k}} \leq a } for \m{t=1,\ldots,N-1,} is solved using multiple shooting methods \cite{karmmsm}. 

\section{Numerical experiments}\label{sec:reswip}

	In this section we perform several numerical experiments on the WIP with parameters lifted from an actual experimental setup. Two key features of our approach will be highlighted  here:
\begin{enumerate}
\item Rotations over \m{360 \si{\degree}}, that would otherwise require the use of multiple charts, are handled in a coherent fashion, and 
\item forward and reverse motions are also handled seemlessly. 
\end{enumerate}

	The following parameters of the WIP system, taken from the prototype that has been developed in the Institute of Automatic
    Control, TU Munich \cite[p.\ 173]{sergiothesis}, were considered for control synthesis:
\begin{table}[H]
\centering
\begin{tabular}{lccc}
\hline
\hline
Model parameters & Symbol & Value & Unit \\
\hline \hline
Body mass & \m{m_b} & 0.277 & \m{\si{\kilogram}} \\
% \hline
Distance from the wheel & \m{b} & \m{48.67\cdot 10^{-3}} & \m{\si{\meter}} \\
axis to the body's center  & & & \\
of gravity  & & & \\
% \hline
Gravitational pull & \m{\mathrm{g}} & \m{9.81} & \m{\si{\meter/\second^2}} \\
% \hline
Wheel mass & \m{m_w} & 0.028 & \m{\si{\kilogram}} \\
% \hline
 Wheel radius  & \m{r_w} & \m{33.1 \cdot 10^{-3}} & \m{\si{\meter}} \\
 % \hline
 Distance between wheels & \m{2d_w} & \m{2 \cdot 49 \cdot 10^{-3}} & \m{\si{\meter}} \\
 %  \hline
 Body's moment of inertia & \m{I_B} & & \\
 \quad around x-axis & \m{I_{Bxx}} & \m{543.108 \cdot 10^{-6}} & \m{\si{\kilogram \meter^2}} \\
\quad around y-axis & \m{I_{Byy}} & \m{481.457 \cdot 10^{-6}} & \m{\si{\kilogram \meter^2}} \\
\quad around z-axis & \m{I_{Bzz}} & \m{153.951 \cdot 10^{-6}} & \m{\si{\kilogram \meter^2}} \\
% \hline
Wheel's moment of inertia & \m{I_W} & & \\
\quad around y-axis (rotation axis) & \m{I_{Wyy}} & \m{7.411 \cdot 10^{-6}} & \m{\si{\kilogram \meter^2}} \\
\quad around z-axis & \m{I_{Wzz}} & \m{4.957 \cdot 10^{-6}} & \m{\si{\kilogram \meter^2}} \\
Control torque bound & \m{\mu} & \m{8 \cdot 10^{-3}}  & \m{\si{\newton\meter\second}} \\
Maneuvers time duration & \m{T} & \m{10-20} & \m{\si{\second}}\\
Sampling time & \m{h} & \m{0.05} &  \m{\si{\second}} \\
\hline
\hline
\end{tabular}
\end{table}

\begin{figure}[H]
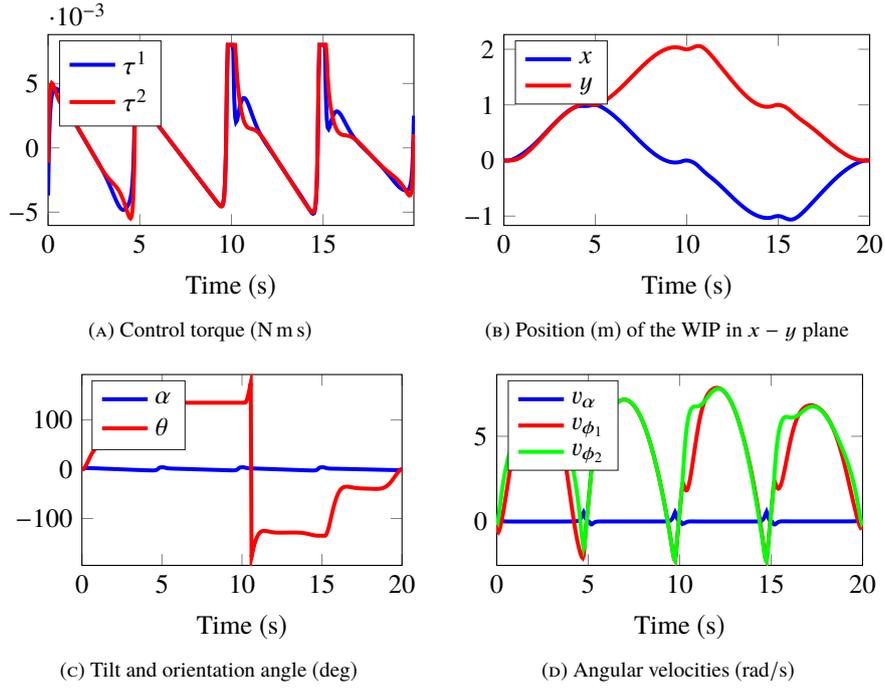

\centering
\subfloat[Control torque \m{\left(\si{\newton\meter\second}\right)}]{
\setlength\figureheight{=0.2\textwidth} 
\setlength\figurewidth{=0.4\textwidth}
\input{OptConMan1.tikz}
\label{fig:OptConMan1}
}
\;
\subfloat[Position \m{\left(\si{\meter}\right)} of the WIP in \m{x-y} plane]{
\setlength\figureheight{=0.2\textwidth} 
\setlength\figurewidth{=0.4\textwidth}
\input{PositionMan1.tikz}
\label{fig:PositionMan1}
}
\;
\subfloat[Tilt and orientation angle \m{\left(\text{deg}\right)}]{
\setlength\figureheight{=0.2\textwidth} 
\setlength\figurewidth{=0.35\textwidth}
\input{TiltOrntMan1.tikz}
\label{fig:TiltOrntMan1}
}
\;
\subfloat[Angular velocities \m{\left(\si{\radian / \second}\right)}]{
\setlength\figureheight{=0.2\textwidth} 
\setlength\figurewidth{=0.4\textwidth}
\input{VelocityMan1.tikz}
\label{fig:VelocityMan1}
}
\caption{Optimal torques and the corresponding state trajectories for the maneuver \m{\mathcal{M}_1}.}
\label{fig:WipMan1}
\end{figure}

\begin{figure}[H]
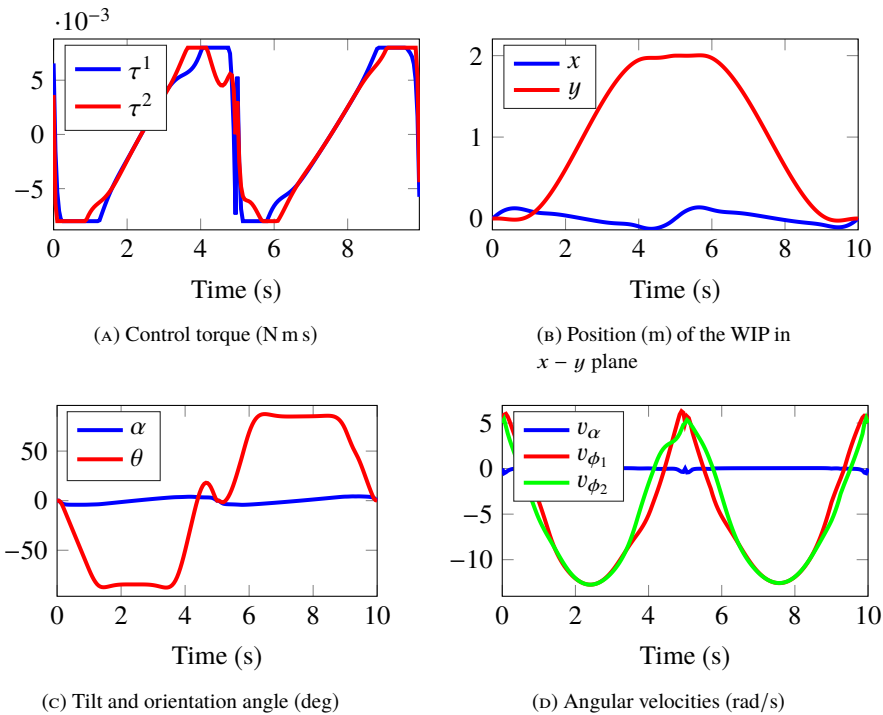

\centering
\subfloat[Control torque \m{\left(\si{\newton\meter\second}\right)}]{
\setlength\figureheight{=0.2\textwidth} 
\setlength\figurewidth{=0.4\textwidth}
\input{OptConMan2.tikz}
\label{fig:OptConMan2}
}
\;
\subfloat[Position \m{\left(\si{\meter}\right)} of the WIP in \m{x-y} plane]{
\setlength\figureheight{=0.2\textwidth} 
\setlength\figurewidth{=0.4\textwidth}
\input{PositionMan2.tikz}
\label{fig:PositionMan2}
}
\;
\subfloat[Tilt and orientation angle \m{\left(\text{deg}\right)}]{
\setlength\figureheight{=0.2\textwidth} 
\setlength\figurewidth{=0.35\textwidth}
\input{TiltOrntMan2.tikz}
\label{fig:TiltOrntMan2}
}
\;
\subfloat[Angular velocities \m{\left(\si{\radian / \second}\right)}]{
\setlength\figureheight{=0.2\textwidth} 
\setlength\figurewidth{=0.4\textwidth}
\input{VelocityMan2.tikz}
\label{fig:VelocityMan2}
}
\caption{Optimal torques and the corresponding state trajectories for the maneuver \m{\mathcal{M}_2}.}
\label{fig:WipMan2}
\end{figure}

The following two constrained maneuvers have been simulated:
\begin{itemize}[leftmargin=*]
\item  \textit{Maneuver 1 \m{\left(\mathcal{M}_1\right):}} For a given set of configurations
\begin{enumerate}
\item \m{a \Let \left(g^a,\alpha^a,v^a \right) = \left(\left(0,0,0 \si{\degree}\right),0\si{\degree},\left(0,0,0 \right) \si{\radian/\second} \right),}
\item \m{b \Let \left(g^b,\alpha^b,v^b \right) = \left(\left(1,1,135 \si{\degree}\right),5\si{\degree},\left(0,1,1 \right) \si{\radian/\second} \right),}
\item \m{c \Let \left(g^c,\alpha^c,v^c \right) = \left(\left(0,2,135 \si{\degree}\right),5\si{\degree},\left(0,0.5,0.5 \right) \si{\radian/\second} \right),}
\item \m{d \Let \left(g^d,\alpha^d,v^d \right) = \left(\left(-1,1,0 \si{\degree}\right),5\si{\degree},\left(0,0.5,0.5 \right) \si{\radian/\second} \right),}
\end{enumerate} 
we find an energy-optimal trajectory that passes through the points \m{a,b,c,d,} traversing each pair in an interval of \m{5\, \si{\second}} in the following order
\[ a \rightsquigarrow b \rightsquigarrow c \rightsquigarrow d \rightsquigarrow a,\]
and satisfies control constraints throughout the journey.

\item  \textit{Maneuver 2 \m{\left(\mathcal{M}_2\right):}} For a given set of configurations
\begin{enumerate}
\item \m{A \Let \left(g^A,\alpha^A,v^A \right) = \left(\left(0,0,0 \si{\degree}\right),0\si{\degree},\left(0,5,5 \right) \si{\radian/\second} \right),}
\item \m{B \Let \left(g^B,\alpha^B,v^B \right) = \left(\left(0,2,0 \si{\degree}\right),0\si{\degree},\left(0,5,5 \right) \si{\radian/\second} \right),}
\end{enumerate} 
we find an energy-optimal trajectory that passes through the points \m{A,B,} in an interval of \m{5\,\si{\second}} in the following order
\[ A \rightsquigarrow B \rightsquigarrow A,\]
and satisfies control constraints throghout the journey.
\end{itemize}
	The optimal control profiles and the corresponding state trajectories for the maneuver \m{\mathcal{M}_1} are given in Figure~\ref{fig:WipMan1}; for \m{\mathcal{M}_2} the corresponding plots are given in Figure~\ref{fig:WipMan2}. 

\begin{figure}[H]
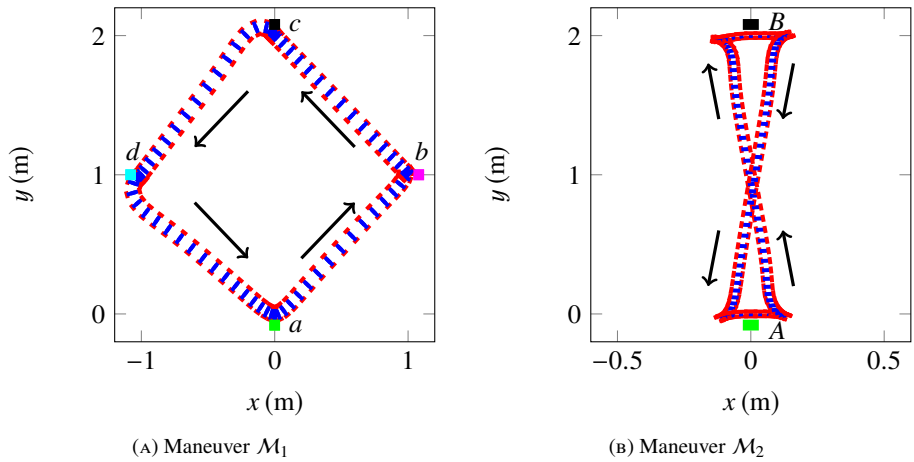

\centering
\subfloat[Maneuver \m{\mathcal{M}_1}]{
\setlength\figureheight{=0.35\textwidth} 
\setlength\figurewidth{=0.35\textwidth}
\input{PortraitMan1.tikz}
\label{fig:PortraitMan1}
}
\;
\subfloat[Maneuver \m{\mathcal{M}_2}]{
\setlength\figureheight{=0.35\textwidth} 
\setlength\figurewidth{=0.35\textwidth}
\input{PortraitMan2.tikz}
\label{fig:PortraitMan2}
}
\caption{Path traced by the WIP on the \m{x-y} plane.}
\label{fig:PortraitWip}
\end{figure} 

	The distinguishing feature of this approach is that the system dynamics is defined on the configuration manifold in contrast to local coordinates, and this enables one to execute maneuvers that need more than one chart. It is evident from Figure~\ref{fig:TiltOrntMan1} that the system takes a \m{360\si{\degree}} turn in order to execute the maneuver \m{\mathcal{M}_1}, and this would require employment multiple charts in other techniques. The optimal controls in Figure~\ref{fig:OptConMan2} saturate for time intervals \m{[3.5\; 5]\, \si{\second}} and \m{[8.5\; 9.5]\, \si{\second}} in order to execute the maneuver \m{\mathcal{M}_2} in the  specified time of \m{10\, \si{\second}}.

	The path traced by the WIP on the \m{x-y} plane for the maneuvers \m{\mathcal{M}_1} and \m{\mathcal{M}_2} are given in Figure~\ref{fig:PortraitWip}. It is important to note that the maneuver \m{\mathcal{M}_2} as shown in Figure~\ref{fig:PortraitMan2} is symmetric about \m{y-}axis because the system admits translational and rotational symmetry and the boundary conditions in the base variables \m{(s,v)} are identical. Another important observation is that the state trajectory from \m{A} to \m{B} in Figure~\ref{fig:PortraitMan2} is traversed by going forward from the point \m{A} due to the initial forward velocity, coming to rest, followed by moving in reverse direction to reach the point \m{B}. This path is chosen to illustrate the fact that the system can traverse in the forward and the backward directions with equal ease under our control synthesis technique. 
   
\bibliographystyle{siam}
\bibliography{refsWIP}
\end{document}